\documentclass[showpacs,superscriptaddress,aps,prb,twocolumn,floats,10pt]{revtex4-1}
\usepackage{graphicx}
\usepackage{amsmath}
\usepackage{amssymb}
\usepackage{epstopdf}
\usepackage{color}
\usepackage{hyperref}

\usepackage[T1]{fontenc}
\usepackage{filecontents}

\newcommand{\e}{\mathrm{e}}

\begin{document}
\title{Magnetic field dependence of the atomic collapse state in graphene}

\author{D.~Moldovan}\email{dean.moldovan@uantwerpen.be}
\affiliation{Department of Physics, University of Antwerp, Groenenborgerlaan 171, B-2020 Antwerpen, Belgium}
\author{M.~Ramezani Masir}\email{mrmphys@gmail.com}
\affiliation{Department of Physics, University of Texas at Austin, Texas 78712, USA}
\author{F.~M.~Peeters}\email{francois.peeters@uantwerpen.be}
\affiliation{Department of Physics, University of Antwerp, Groenenborgerlaan 171, B-2020 Antwerpen, Belgium}

\begin{abstract}
Quantum electrodynamics predicts that heavy atoms ($Z > Z_c \approx 170$) will undergo the process of \textit{atomic collapse} where electrons sink into the positron continuum and a new family of so-called \textit{collapsing} states emerges.
The relativistic electrons in graphene exhibit the same physics but at a much lower critical charge ($Z_c \approx 1$) which has made it possible to confirm this phenomenon experimentally.
However, there exist conflicting predictions on the effect of a magnetic field on atomic collapse.
These theoretical predictions are based on the continuum Dirac-Weyl equation, which does not have an exact analytical solution for the interplay of a supercritical Coulomb potential and the magnetic field.
Approximative solutions have been proposed, but because the two effects compete on similar energy scales, the theoretical treatment varies depending on the regime which is being considered.
These limitations are overcome here by starting from a tight-binding approach and computing exact numerical results.
By avoiding special limit cases, we found a smooth evolution between the different regimes.
We predict that the atomic collapse effect persists even after the magnetic field is activated and that the critical charge remains unchanged.
We show that the atomic collapse regime is characterized: 1) by a series of Landau level anticrossings and 2) by the absence of $\sqrt{B}$ scaling of the Landau levels with regard to magnetic field strength.
\end{abstract}


\maketitle

\twocolumngrid

\section{Introduction}

Even before its experimental isolation in 2004~\cite{Novoselov2004}, graphene was considered as an analog of ($2 + 1$) dimensional quantum electrodynamics \cite{Semenoff1984,DiVincenzo1984,Haldane1988}.
Its two-dimensional crystal lattice hosts massless Dirac fermions which move with a Fermi velocity of about 1/300 the speed of light with a linear spectrum close to the $K$ and $K'$ points of the Brillouin zone.\cite{CastroNeto2009}
Examples of its relativistic properties include the Klein paradox~\cite{Klein1929,Katsnelson2006} and Zitterbewegung~\cite{itzykson2006}.
The detection of the anomalous integer quantum Hall effect served as the definitive demonstration of the relativistic nature of its carriers as well as the signature of zero-gap single-layer graphene.\cite{Novoselov2005,Zhang2005}

It was shown that graphene exhibits the analogue of atomic collapse \cite{Pereira2007,Shytov2007,Shytov2007a}, a fundamental phenomena in quantum electrodynamics (QED).
When the charge of an atomic nucleus exceeds a certain critical value, the energy levels of the bound electronic states dive into the lower positron continuum and the spontaneous generation of electron-positron pairs is expected.\cite{Pomeranchuk1945,Zeldovich1972,Greiner1985a}
The extremely high charge requirements ($Z_c \approx 170$) have prevented the observation of this phenomenon with real atoms.
However, thanks to the relativistic nature of the carriers, no band gap and the large effective fine structure constant in graphene, the same physics can be observed at a much lower charge ($Z_c \approx 1$).
The switch to graphene also changes the energy scale from MeV to sub-eV and the spontaneous pair creation changes from electron-positron to electron-hole.
This has made it possible to realize this phenomena experimentally, with observations closely matching the predictions of QED.\cite{Wang2013,Mao2016,Jiang2017}

Another longstanding prediction from QED is that a magnetic field should be able to enhance the effect of collapse.\cite{Oraevski1978}
A magnetic field confines the motion of the electron, therefore bringing it closer to the nucleus.
As a result, the required value of the critical charge decreases as a function of the field strength.
However, this is where the graphene analogue may diverge from the original.
QED considers (3 + 1) dimensions where the magnetic field acts on the electron in a plane, but not on the other degree of freedom.
Due to its flat nature, the electrons in graphene are confined to (2 + 1) dimensions, which results in a different problem, e.g. the 2D electron energy is completely quantized into Landau levels.

It was shown experimentally that a charged impurity in graphene lifts the orbital degeneracy of Landau levels, thus splitting them into discrete states.\cite{Luican-mayer2014}
However, this experiment was limited to a charge in the subcritical regime.
Previous theoretical studies have had conflicting conclusions about the influence of the magnetic field on the value of the critical charge.
For the problem of a magnetic monopole, Ref.~\onlinecite{Valenzuela2016} predicts the absence of atomic collapse.
In Ref.~\onlinecite{Gamayun2011}, it was predicted that the critical charge vanishes for massless carriers at any finite magnetic field, making any charge supercritical in gapless graphene.
This was further explored in Ref.~\onlinecite{Sobol2016}.
On the other hand, Refs.~\onlinecite{Zhang2012,Maier2012,Kim2014} argued that the critical charge will not change.
It was also shown that the effect persists in the discrete energy spectrum of a quantum dot\cite{VanPottelberge2017}, which further supports the argument that quantization of the energy does not influence the critical charge.

The problem of a supercritical charge in a magnetic field cannot be solved exactly in analytic form and the non-perturbative nature of the addition of a magnetic field makes the problem complicated.
Both electric and magnetic fields compete on similar energy scales and the applied approximations may vary depending on the regime which is being studied (subcritical vs. supercritical charge, zero vs. finite magnetic field).
The different theoretical treatment of the regimes makes it difficult to see the transitions from one to another.
This can also highlight the differences rather than the similarities of the regimes.

In this paper, we investigate the problem of a supercritically charged impurity in graphene in the presence of a magnetic field using the numerical tight-binding approach.
Compared to previous work, this approach allows us to treat all regimes equally: from zero to finite fields, including unbalanced or equal contributions of both the electric and magnetic fields.
We compute the evolution between regimes with numerically exact results and we find that the atomic collapse effect persists even after the magnetic field is activated and that the critical charge remains unchanged.
We show that the atomic collapse resonances are directly connected with a series of Landau level anticrossings.
These avoided crossings are formed by low-energy orbital states which split off from the unperturbed Landau levels.
They are accompanied by a strong enhancement of the LDOS close to the impurity, similarly to the zero-magnetic-field case.
Another aspect is the $\sqrt{B}$ scaling of the Landau levels which is broken for certain levels in the supercritical regime.
By computing the evolution of the system with magnetic field strength, we show a direct correspondence between the atomic collapse resonance and anomalous scaling of the Landau levels.

\section{Theoretical Model}

The tight-binding Hamiltonian for graphene in the presence of a charged impurity is given by\cite{CastroNeto2009}
\begin{equation}\label{hamiltonian_b}
\begin{split}
    H =& \sum_{\left< i,j \right>} \left( t_{ij} a_i^\dagger b_j + H.c. \right)
     \\* &+ \sum_i V(\vec{r}_i^{\ A}) a_i^\dagger a_i
     \\* &+ \sum_i V(\vec{r}_i^{\ B}) b_i^\dagger b_i,
\end{split}
\end{equation}
where $t_{ij} = -2.8$ eV is the hopping energy between lattice sites $i$ and $j$, operators $a_i (a_i^\dagger)$ and $b_i (b_i^\dagger)$ create (annihilate) an electron at site $i$ of sublattice $A$ and $B$, respectively, and $\vec{r}_i^{\ A,B}$ is the position of the carbon atoms relative to the impurity.
The first term includes the interaction of the nearest-neighbor hopping pairs $\left< i,j \right>$ in graphene (next-nearest and higher hoppings do not have an impact on the results of this paper).
The last two terms include the potential on the atoms of each sublattice.
A Coulomb center of charge $Z$ generates the potential $V(r) = -\hbar v_{F} \beta / r$ where $\beta \equiv Z e^2 / \kappa \hbar v_{F}$ is the dimensionless coupling constant.
The raw impurity charge $Z$ is scaled by the relative permittivity $\kappa$ and the Fermi velocity $v_F$.
For convenience, we mainly use $\beta$ to denote the charge of the impurity.

By applying the low-energy approximation, Eq.~\eqref{hamiltonian_b} can be simplified to reveal the Dirac equation which governs states near the $K$ points,
\begin{equation}\label{dirac_eq_v}
    \hbar v_F \left(-\mathrm{i} \vec{\sigma} \cdot \vec{\nabla} - \frac{\beta}{r} \right) \Psi(\vec{r}) = E \Psi(\vec{r}),
\end{equation}
where $\vec{\sigma} = (\sigma_x, \sigma_y)$ are the Pauli matrices.
Due to the axial symmetry of the potential, Eq.~\eqref{dirac_eq_v} is separable in cylindrical coordinates using the relation $x \pm i y = r e^{\pm i\varphi}$ and the solutions can be found in the form \cite{DiVincenzo1984},
\begin{equation}
    \Psi_m(r) = \frac{1}{\sqrt{r}}
    \begin{bmatrix}
        \e^{\mathrm{i} m\varphi} a_m(r) \\
        \mathrm{i} \e^{\mathrm{i} (m + 1)\varphi} b_m(r)
    \end{bmatrix},
\end{equation}
where $m = 0, \pm 1, \pm 2, ...$ is the orbital quantum number. Equation~\eqref{dirac_eq_v} therefore reduces to,
\begin{equation}
    \begin{bmatrix}
        \frac{E}{\hbar v_F} + \frac{\beta}{r}  & -\partial_r - \frac{m+1}{r} \\
        \partial_r - \frac{m}{r}       & \frac{E}{\hbar v_F} + \frac{\beta}{r}
    \end{bmatrix}
    \begin{bmatrix}
    a_m(r) \\ b_m(r)
    \end{bmatrix}
    = 0.
\end{equation}
This coupled pair of first order differential equations can be reduced to two decoupled second order equations.
In the limit $r \to 0$, the solution behaves as\cite{Pereira2007,Shytov2007}
\begin{equation}\label{psi0}
    \varphi_m(r) \sim r^\gamma, \quad \gamma = \sqrt{(m+1/2)^2 - \beta^2}.
\end{equation}

This reveals a problem for the lowest angular momentum modes ($m=-1, 0$), because $\gamma$ becomes imaginary if $\beta > \beta_c = 1/2$.
In this case the solution oscillates endlessly towards the center as $\e^{i\log{r}}$.
From a classical perspective this can be understood as a critical angular momentum above which the orbits spiral and fall into the potential origin.\cite{Shytov2007a}
Thus, a charge in graphene which exceeds the critical value $\beta_c$ is seen as a supercritical nucleus which triggers the analogue of the atomic collapse phenomena.

As we have just shown, the quantum-mechanical problem is ill-defined for a point charge in the supercritical regime ($\beta > 1/2$).
An additional boundary condition must be introduced to cut off the potential at short distances.
This is analogous to the introduction of the finite size of the nucleus in QED.\cite{Zeldovich1972}
In graphene experiments, this is defined by the radius of the artificial nucleus used to build up a supercritical charge, e.g. a cluster of Ca dimers\cite{Wang2013} or a charged vacancy\cite{Mao2016}.
The modified impurity potential with cutoff radius $r_0$ reads,
\begin{equation}\label{cutoff_potential}
  V_\beta(r) =
  \begin{cases}
      -\hbar v_F \frac{\beta}{r_0}, & \text{if}\ r \le r_0\\
      -\hbar v_F \frac{\beta}{r}, & \text{if}\ r > r_0
  \end{cases}.
\end{equation}
The choice of cutoff length $r_0$ will be discussed in the next section.
With the introduction of this more realistic potential, the eigenvalue problem becomes well defined even for $\beta > 1/2$, however it can no longer be solved in analytical form and instead requires a numerical treatment.

In addition to the finite-sized supercritical charge, a homogeneous magnetic field $B$ is included via the vector potential $\vec{A} = B/2 (y, -x)$ in the symmetric gauge.
The Dirac equation thus reads,
\begin{equation}\label{dirac_eq_b}
    \left(v_F \vec{\sigma} \cdot (-\mathrm{i}\hbar\vec{\nabla} + e\vec{A}) + V_\beta(\vec{r}) \right) \Psi(\vec{r}) = E \Psi(\vec{r}).
\end{equation}
The wavefunction components $a_m(r)$ and $b_m(r)$ must satisfy the following coupled first-order differential equations:
\begin{equation}
    \begin{bmatrix}
        \frac{E - V_\beta(r)}{\hbar v_F}  & -\partial_r - \frac{m+1}{r} + \frac{r}{2l_B^2} \\
        \partial_r - \frac{m}{r} + \frac{r}{2l_B^2}       & \frac{E - V_\beta(r)}{\hbar v_F}
    \end{bmatrix}
    \begin{bmatrix}
    a_m(r) \\ b_m(r)
    \end{bmatrix}
    = 0.
\end{equation}
Solving these equations analytically, including both the Coulomb potential and the homogeneous magnetic field, is not possible except for special limit cases.
Taking the limit $r \rightarrow 0$ actually recovers the same result as before, in Eq.~\eqref{psi0}, which neglects the influence of the magnetic field.
On the other hand, taking $r \rightarrow \infty$ yields an unperturbed Landau level sequence, free of the influence of the charge.
However, we are most interested in the situation at moderate $r$ where the effects of the charge and magnetic field both feature prominently.

In the absence of the impurity ($\beta=0$), the eigenvalue problem can be solved analytically and yields the Landau level (LL) sequence with energy $E_N = \pm \hbar v_F / l_B \sqrt{2|N|}$, where $l_B = \sqrt{\hbar/(eB)}$ is the magnetic length, $N = 0, \pm1, \pm2, ...$ is the level index, and +(-) refers to electron (hole) states.
Pristine graphene is translationally invariant and so LL energy $E_N$ is independent of $m$.
Each Landau level $N$ is degenerate, consisting of an infinite number of states with wavefunctions $\Psi_{Nm}(r)$ where the orbital number $m \geq -|N|$.\cite{Goerbig2011}

As has been shown experimentally in Ref.~\onlinecite{Luican-mayer2014}, adding an impurity ($\beta > 0$) breaks the translational symmetry, thus lifting LL degeneracy.
The energy splits into $m$-dependent sublevels $E_{Nm}$.
The lowest energy orbital states are centered around the impurity, while higher order states form concentric orbits around it.
However, the impurity available in the experiment was only within the subcritical regime.
The theoretical study in Ref.~\onlinecite{Sobol2016} incorporated a supercritical charge, but concluded that the presence of a finite magnetic field presents a distinctly different regime as compared to the zero-field atomic collapse phenomena in graphene.

\begin{figure*}[t]
    \includegraphics[width=\textwidth]{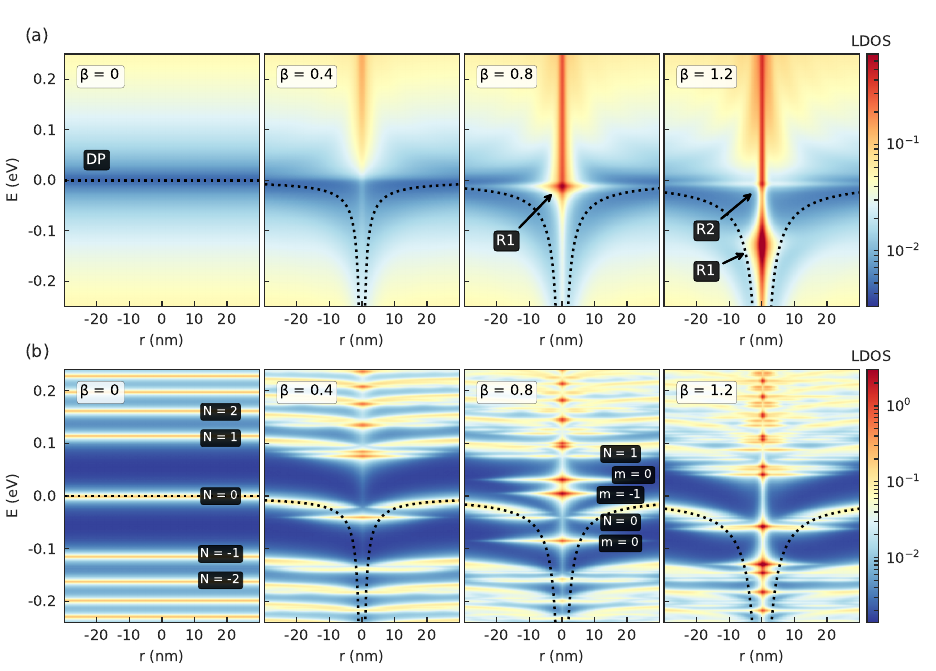}
    \caption{Colormap of the LDOS as a function of position (relative to the impurity) and energy. (a) The top row presents the LDOS in the absence of a magnetic field ($B=0$), for a charge $\beta$ from $0$ to $1.2$ as indicated in the panels. The dotted line shows the position of the Dirac point (DP) in the first panel and the spatial profile of the Coulomb potential in the following panels. The collapse resonances are labeled $R1$ and $R2$.
    (b) The bottom row presents the results with a magnetic field of $B=12$ T. The Landau levels are labeled as $N=0, \pm1, \pm2$, while the $m$ labels indicate the split orbital states.}
    \label{fig:ldos_r}
\end{figure*}

\begin{figure*}[t]
    \includegraphics[width=\textwidth]{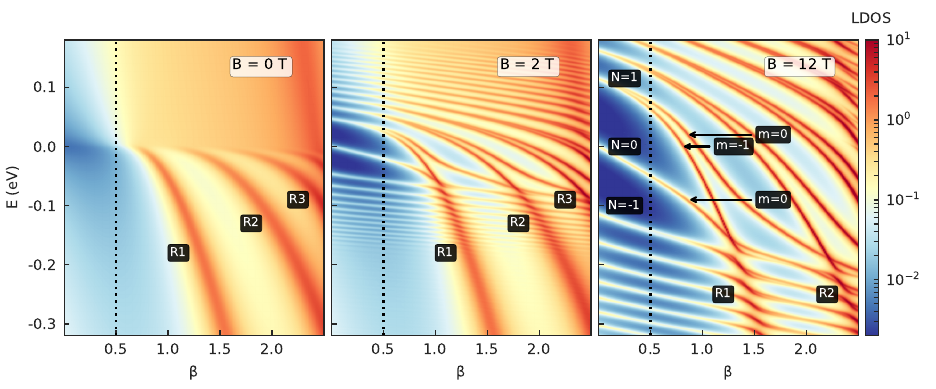}
    \caption{Colormap of the LDOS at the impurity ($r=0$) as a function of charge $\beta$ and energy $E$. The magnetic field $B$ is indicated in the panels. The vertical dotted line marks $\beta_c = 1/2$. The labels $R1$ to $R3$ indicate the collapse resonances in the order in which they appear. The $N=0, \pm1$ labels indicate the different Landau levels and orbital states $m$.}
    \label{fig:LDOS_beta}
\end{figure*}

\section{Numerical tight-binding approach}

We consider the full tight-binding Hamiltonian as given in Eq.~\eqref{hamiltonian_b}.
We take the electric potential profile of the impurity $V(r)$ as given by Eq.~\eqref{cutoff_potential}.
The cutoff radius $r_0$ accounts for the finite size of the charge and is usually taken to match the size of the impurity.
Here, we take $r_0 = 0.5$ nm which is in line with experimental data.~\cite{Wang2013,Mao2016}
In experiments, a constant charge $Z$ may be present, while the relevant coupling $\beta \equiv Z e^2 / \kappa \hbar v_{F}$ may be tuned by applying a gate voltage which controls the relative permittivity $\kappa$ via Landau level occupancy (i.e. screening of the charge $Ze$).~\cite{Luican-mayer2014}
In the theoretical model we change the charge coupling directly through the parameter $\beta$.
In the presence of a uniform magnetic field of strength $B$, perpendicular to the graphene plane, the hopping parameters are replaced by the Peierls substitution, $t_{ij} \rightarrow t_{ij} \e^{i 2\pi \Phi_{ij} }$, where $\Phi_{ij} = (1/\Phi_0) \int_{\vec{r}_i}^{\vec{R_j}} \vec{A} \cdot \vec{dl}$ is the Peierls phase and $\Phi_0 = h/e$ the magnetic quantum flux.

For the computation, we construct a large finite-sized model system in the shape of a hexagonal flake with armchair edges (in order to avoid zigzag edge states with low energy).
The impurity is positioned in the center and the flake is taken sufficiently large such that its finite size does not influence the physics we are interested in.
In the following calculations we take the hexagon edge width of $200$~nm, which corresponds with a flake consisting of about four million carbon atoms.
The model system is built using our open-source code package for numerical tight-binding calculations: \textit{pybinding}.\cite{pybinding094}
The package includes a fast implementation of the kernel polynomial method~\cite{Weisse2006,Covaci2010} which is used for the calculation of the local density of states in this paper.

\section{Results}

Before turning on the magnetic field, we shall briefly review the real-space picture of the atomic collapse resonances in the local density of states (LDOS) in graphene.
The LDOS can be measured using scanning tunneling spectroscopy (STS) and provides an experimentally observable signature of atomic collapse as shown in Refs.~\onlinecite{Wang2013,Mao2016,Jiang2017}.
Figure \ref{fig:ldos_r}(a) presents the space-energy map of the LDOS in the subcritical and supercritical regimes.
A subcritical charge ($\beta = 0.4$) enhances the LDOS in the positive part of the spectrum in proximity of the impurity ($r = 0$).
Note that it does not cross below the Dirac point.
Once the charge becomes supercritical ($\beta = 0.8 > \beta_c$) the high intensity LDOS region crosses below zero energy.
The atomic collapse states can only be found at negative energy since they represent coupled states, where an electron from the center can tunnel out and escape as a hole.
The collapse resonance in the LDOS is labeled $R1$ as the first of such states to appear with increasing charge $\beta$.
The LDOS intensity is highest at the center, but disappears quickly at about $10$ nm away from the impurity.

\subsection*{Level splitting}

The result of the LDOS computation for a magnetic field of $B=12$ T are presented in Fig.~\ref{fig:ldos_r}(b).
Without the impurity ($\beta = 0$), the Landau levels appear constant in space, as expected.
When a small charge is introduced ($\beta = 0.4$) the Landau levels start to bend and split into individual orbital states near the impurity ($r=0$).
When the charge is increased ($\beta = 0.8$), multiple split levels are clearly visible.
States with smaller orbital numbers have lower energy and sink down with the Coulomb potential.
At LL $N=0$, the orbital state $m=0$ is clearly separated.
Similarly, at LL $N=1$ states $m=-1$ and $m=0$ have moved lower and separate from the rest of the LL.
Notice that the LDOS at the impurity is an order of magnitude higher than the surrounding area, similar to the case without a magnetic field.
After further increasing the charge to $\beta = 1.2$, many more states split off and dive to low energy.
Keeping track of the many states becomes difficult and therefore we shall switch to the evolution of the LDOS$(r = 0)$ with $\beta$ in the next section.

\subsection*{Anticrossing series}

The resonances associated with atomic collapse in graphene have high intensity only in close proximity to the charge.
They disappear quickly, only a few nanometers away from the impurity.
For this reason, we will be focusing mainly on the properties at the point of the impurity ($r=0$).
We calculated the LDOS as a function of energy $E$ and the charge of the impurity $\beta$ in Fig.~\ref{fig:LDOS_beta} for various magnetic fields $B$.

Without a magnetic field, the signature of collapse is easy to spot as high intensity resonances at negative energy (labeled $R1$-$R3$ in Fig.~\ref{fig:LDOS_beta}).
As $\beta$ is increased the $R1$ resonance moves down and broadens, while a second ($R2$) resonance appears just below the Dirac point for $\beta > 1$.
Both resonances are clearly set apart from the rest of the (mostly homogeneous) local density of states.

When a magnetic field is applied ($B=2$ T and $12$ T in Fig.~\ref{fig:LDOS_beta}) we can see a mix of Landau levels and collapse resonances.
Landau levels are clearly formed at low $\beta$.
As the charge $\beta$ is increased, we can see Landau level $N=1$ splitting into individual orbital states $m=-1$ and $m=0$.
At higher $\beta$, Landau level $N=2$ splits, then $N=3$ and so on.

The lower split state ($m=-1$) of each positive LL is of special interest because of their connection with the collapse resonances at $B=0$.
As $\beta$ is increased, level $N=1$, $m=-1$ crosses $N=0$, $m=0$ but is then repelled by $N=-1$, $m=-1$ forming an anticrossing.
This avoided crossing formation was shown previously in Ref.~\onlinecite{Sobol2016} and appears due to the repulsion of levels with the same orbital number $m$.
Here, we show that the anticrossings form a long series which follows the same line as the $R1$ collapse resonance at $B=0$, indicating a clear persistence of the collapse resonance in the presence of the magnetic field.
This line of anticrossings also retains a very high LDOS intensity which is at least an order of magnitude larger than the unaffected Landau levels.

In Ref.~\onlinecite{Gamayun2011}, the critical charge in a magnetic field was defined simply as the point where an electron state with $m=0$ moves below zero energy.
Based on that, it was found that any finite charge is supercritical in a magnetic field in gapless graphene.
However, this criterion is overly broad.
Here, we find that indeed there is a state which immediately sinks to negative energy for $\beta > 0$: this is the $N = 0$, $m=0$ Landau level as seen in Fig.~\ref{fig:LDOS_beta}.
States which split off from the zeroth Landau level are very important in the many body problem, e.g. with regard to the excitonic condensate and magnetic catalysis.\cite{Semenoff1999}
However, in connection with the atomic collapse phenomena, apart from the downshift in energy, we do not see any changes for these states within the subcritical regime, i.e. for $\beta < \beta_c = 0.5$.
New features in the LDOS are only visible when the charge is above $\beta_c$, as described above (and further supported by the breaking of $\sqrt{B}$ scaling in later sections).

There is a striking resemblance between our Fig.~\ref{fig:LDOS_beta} for $B = 12$ T and Fig. 1 in Ref.~\onlinecite{Sobol2016} up to states with $m=0$, but we draw different conclusions.
Reference~\onlinecite{Sobol2016} concludes that a charge in any finite magnetic field presents a distinctly different regime as compared to  zero-field.
We propose that this interpretation is actually a consequence of the different theoretical treatment of the $B=0$ and $B \neq 0$ regimes which makes it impossible to see the transitions from one to another, thus highlighting the differences rather than the similarities of the two regimes.
Our results clearly show a persistence of the same effect, especially in Fig.~\ref{fig:LDOS_beta} for $B=2$ T which very closely resembles the zero-field case.
The $B=0$ and $B=2$ T cases are actually indistinguishable for negative energy states where the LL spacing is very small.

The inter-level spacing of the LLs is generally preserved while $\beta$ changes.
It is only disturbed while crossing a collapse resonance.
After crossing the $R1$ resonance, the negative LLs shift down by one, e.g. $N=-1$ moves lower to take the place of $N=-2$ while $N=-2$ shifts to $N=-3$, etc.
The positive LLs behave differently.
As described earlier, the lower energy state LL ($N=1$, $m=-1$) follows the collapse resonance until it crosses $N=0$. The higher orbital state ($N=1$, $m=0$) continues to the second resonance.
Only levels up to $N=1$ are affected by the $R1$ resonance.
As the charge increases, $N=2$ is included and the pattern repeats with the lowest orbital state diving to low energy and starting the anticrossing series of resonance $R2$.

\begin{figure*}[t]
    \includegraphics[width=\textwidth]{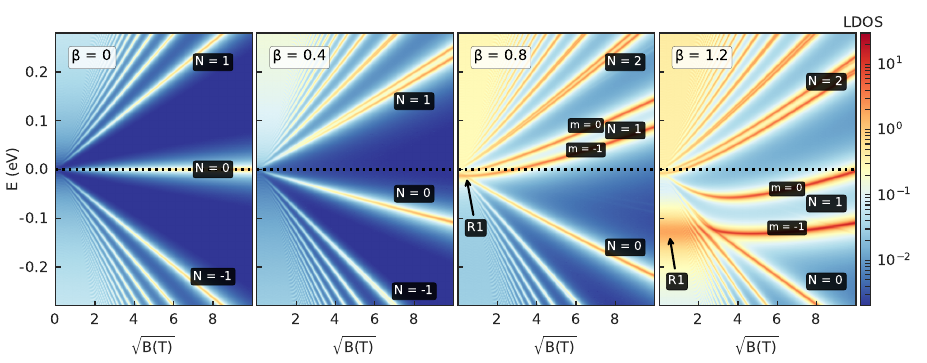}
    \caption{Colormap of the LDOS at the impurity ($r=0$) as a function of energy $E$ and the square root of the magnetic field $\sqrt{B}$. The charge $\beta$ varies as labeled in the panels.
    The first panel ($\beta=0$) shows regular Landau levels with labels up to $N=\pm1$. In the last panel ($\beta=1.2$) the individual orbital states of LL $N=1$ are indicated as $m=-1$ and $m=0$.
    }
    \label{fig:LDOS_B1}
\end{figure*}

Figure~\ref{fig:LDOS_beta} presents the LDOS at the impurity ($r=0$) where only the lowest angular momentum states contribute to the anticrossings.
On the other hand, regular crossings appear as a mix of low and high angular momentum states.
This can be seen in Fig.~\ref{fig:ldos_r}(b) for $\beta = 1.2$ where the higher $m$ states appear away from $r=0$.
The lowest angular momentum states of the $N > 0$ LLs (electron states within the impurity potential) cross the higher angular states of $N \le 0$ LLs (hole states away from the charge).
This is the same kind of electron-hole pairing that existing in the zero-magnetic-field case.

By comparing the results for $B=2$ T and $B=12$ T in Fig.~\ref{fig:LDOS_beta}, notice that the increased magnetic field strength causes the first anticrossing to appear at higher $\beta$.
The anticrossings still appear on the same line as the zero-field resonance, but the energy spacing increases as a consequence of the stronger field.
However, this does not indicate that the value of $\beta_c$ depends on the magnetic field.
While the anticrossings are a feature of the supercritical regime, they are discrete and cannot show a complete picture of atomic collapse in a magnetic field.
Anomalies appear even between these crossings, as we shall show in the next two sections.

\subsection*{Landau level bending}

The Landau levels in graphene feature a linear dependence on the square root of the magnetic field $\sqrt{B}$ as a consequence of the linear energy spectrum.
Computing the LDOS as a function of $\sqrt{B}$ without any charge ($\beta = 0$) reveals the expected linear LL lines of high LDOS intensity in Fig.~\ref{fig:LDOS_B1}.
When a small charge is added ($\beta = 0.4$), the LLs remain linear, but they are slightly tilted downwards.
At this point LL $N=1$ is split into individual orbital states ($m=-1$ and $m=0$).
As the charge is increased into the supercritical regime ($\beta = 0.8$) the levels become non-linear.
The collapse resonance $R1$ is visible near $\sqrt{B} = 0$.

The curvature of the Landau levels becomes especially visible for $\beta = 1.2$.
The $R1$ resonance is quite apparent at low values of $\sqrt{B}$.
Without a magnetic field, this peak appears as a broad LDOS resonance which is an indicator of the supercritical regime.
Notice that the $R1$ resonance (low $\sqrt{B}$) is located at the same energy as the $m=-1$ state of Landau level $N=1$ (high $\sqrt{B}$), which further supports the connection of the collapse resonance and the lowest orbital LL states.

The LLs also appear to bend in the energy region affected by the collapse resonance, while the unaffected levels remain linear as function of $\sqrt{B}$. The collapse resonance looks to be directly connected to the non-linear scaling of Landau levels.

\subsection*{Scaling anomaly}

\begin{figure}[t]
    \centering
    \includegraphics[width=\columnwidth]{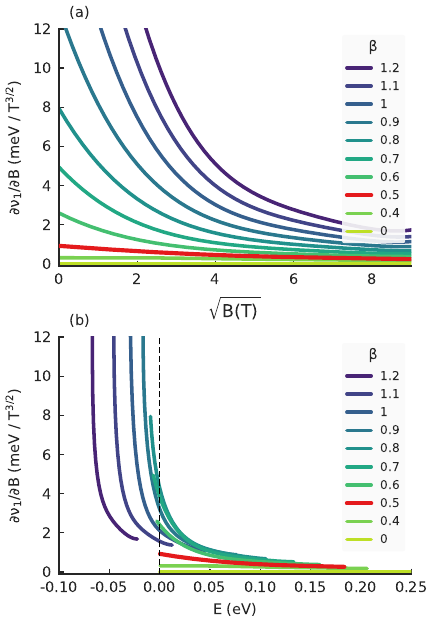}
    \caption{(a) The derivative of the scaling factor $\nu_1$ for different values of the charge $\beta$ for Landau level $N=1$, $m=-1$.
    (b)~Same derivative, but presented as a function of energy instead of the magnetic field.}
    \label{scaling}
\end{figure}

The energy of the Landau level $N$ can be written as,
\begin{equation}
    E_N(B) = \pm v_F \sqrt{2|N| \hbar} \sqrt{B} = \pm \nu_N \sqrt{B},
\end{equation}
where $\nu_N$ is the level scaling factor. When there is no impurity in the system, this factor is constant and independent of the magnetic field, i.e.
\begin{equation} \label{scaling_diff}
    \frac{\partial \nu_N}{\partial B} = 0.
\end{equation}
On the other hand, when Eq.~\eqref{scaling_diff} is non-zero, it means that the level has a scaling anomaly, which should be the case for a supercritical charge.

We use the LDOS data to calculate the derivative $\partial \nu_1 / \partial B$ for LL $N=1$, $m=-1$ and present the results in Fig.~\ref{scaling}(a).
The derivative is close to zero for $\beta$ up to $\beta_c = 0.5$, independent of the magnetic field.
For $\beta > 0.5$ there is a clear non-zero derivative at small values of the magnetic field.
As the magnetic field is increased, the derivative approaches zero asymptotically.
Note that this does not indicate that the scaling anomaly disappears at very high values of the magnetic field.
The scaling anomaly is mainly a function of energy.

In order to highlight the strong energy dependence of the derivative, we change the presentation of the data as shown in Fig.~\ref{scaling}(b).
Notice that the $\beta$ values up to $0.5$ (that have zero derivative, i.e. normal scaling) are located in the region of positive energy.
Once the LL starts crossing into negative energy ($\beta = 0.6$ and higher), the derivative becomes finite, indicating anomalous scaling.
This mirrors the appearance of the collapse resonance below the Dirac point without a magnetic field.
As $\beta$ is increased the LL moves lower in energy (just like the resonance) and the derivative increases indicating stronger anomalous scaling as a function of increasing $\beta$.

\section{Conclusions}

We used the tight-binding method to model a supercritically charged impurity in graphene in the presence of a magnetic field.
We calculated the LDOS in the proximity of the charge as can be observed experimentally using STS.
Without magnetic field, the signature of atomic collapse in graphene is a resonance in the LDOS that forms just below the Dirac point.
When a B-field is present, the resonances are replaced by a Landau level anticrossing series at the same energy, indicating that the value of the critical charge is not affected by the magnetic field.

A Coulomb-like charge causes Landau levels to split into individual orbital states where the lowest ones are of special interest.
When expressed as a function of the charge, these states follow the evolution of the supercritical resonance until they cross the zeroth Landau level.
At that point, a series of anticrossings is formed which continues along the line of the collapse resonance.
The affected Landau level also exhibit an LDOS intensity which is at least an order of magnitude larger than the unaffected levels, which should be observable by STS.

Landau levels which are caught in the collapse resonance also exhibit anomalous scaling as a function of the magnetic field.
The effect closely mirrors the collapse resonance: anomalous scaling appears just as a Landau level crosses below the Dirac point.
An experimental investigation of this phenomenon would require the measurement of LL energy for several magnetic field strengths in order to show the absence of $\sqrt{B}$ scaling in the presence of a supercritical charge.

\section{Acknowledgment}

We thank Eva Andrei, Jinhai Mao and Yuhang Jiang for insightful discussions.
This work was supported by the Flemish Science Foundation (FWO-Vl) and the Methusalem Funding of the Flemish Government.

\end{document}